\documentstyle[aps,preprint]{revtex}
\begin{document}
\title{
Bistability of Persistent Currents in Mesoscopic Rings}
\author{E. V. Anda}
\address{Instituto de F\'{\i}sica,\\
Universidade Federal Fluminense, Av Litoranea s/n, \\
Gragoata, Niteroi, Rio de Janeiro, Brazil}

\author{V. Ferrari and G. Chiappe}
\address{Departamento de F\'\i sica,\\
FCEYN-Universidad de Buenos Aires, \\
c.p. 1428-Nu\~nez, Buenos Aires, Argentina}

\begin{abstract}
We study the persistent currents flowing in a mesoscopic ring threaded 
by a magnetic flux and connected to a stub of finite length.
Multistability processes and Coulomb blockade are demonstrated to be 
present in this system. These properties are functions of the
magnetic flux crossing the ring which plays the role that the
external applied potential fulfills in the multistability behaviour of 
the standard mesoscopic heterostructures.
\end{abstract}
\vskip2cm


\section{Introduction}
The existence of persistent currents in a conducting mesoscopic ring
threaded by a magnetic flux was proposed for the first time by B\"uttiker,
Imry and Landauer \cite{1} and experimentally observed in an ensemble of
rings and in semiconductor and metallic loops \cite{2}. In a ring, the
magnetic flux can be introduced through a boundary condition for the
electronic wave function.
\begin{equation}
\Phi( x + 2\pi r)= \Phi(x) e^{2\pi i \phi/\phi_{0}} 
\end{equation}

where $r$ is the radius of the ring , $\phi _0=h/e$ is a quantum of magnetic
flux and $\phi $ is the real magnetic flux encircled by the loop. Equation
(1) introduces a periodicity to all the physical properties of the system as
a function of $\phi $ with the period associated to $\phi _0$. For a
particular energy level the persistent current flowing along the system can
be obtained by calculating the derivative of the energy with respect to the
magnetic flux. The total current is obtained summing the contributions over
all the occupied states below the Fermi level. 

Since the original experiment \cite{2} there has been a considerable 
theoretical effort to understand currents and
current fluctuations of non-interacting electrons in open and closed rings 
\cite{3,4}. Many works was devoted to study the
effect of electronic interaction upon the current in closed rings 
and the interplay  between correlation and disorder has been reported in this
systems using a number of different methods \cite{4}. From an experimental point of 
view, works emphasizing on different geometries in open and closed systems
has been reported and  transport through a quantum dot embbeded in the ring 
has been studied \cite{5} .
 
We study the persistent currents flowing in a mesoscopic ring
threaded by a magnetic flux and connected  to a stub of finite
length. A change in the magnetic field produces level crossings and 
eventually an interchange of position between
the ground state and the first excited level. This crossing 
 gives
rise to metastable situations. A metastable state   
corresponds to a local minimum of the energy in the parameter space separated 
by an energy barrier from the absolute minimum. So, if this particular
state is obtained through an adiabatic variation of the magnetic flux, 
the system is unable to reach its ground state remaining in the relative local minimum.

This multistability can occur in a interacting perfect ring and also 
in a inhomogeneous one, as it is the case of a perfect ring plus a stub of finite length. In this last case the stub acts as a reservoir of particles and the magnetic flux crossing the ring plays the role that the external applied 
potential fulfills in transport in standard mesoscopic heterostructures.
This system has already been studied \cite{6}
focusing other properties, within the context of the electrochemical
capacity ideas. The structure is described by a tight-binding Hamiltonian
given by:

\begin{equation}
\hat H_{r}=-t \sum_{\sigma,i=1}^{N_{r}} \hat c_{i,\sigma}^{\dag} \hat
c_{i+1,\sigma}+ \sum_{\sigma,\sigma^{^{\prime}},j,i=1}^{N_{r}} U_{i,j}^{
\sigma,\sigma^{^{\prime}}}\hat
n_{i,\sigma}^{c} \hat n_{j,\sigma^{^{\prime}}}^{c} \nonumber
\end{equation}
\begin{equation}
\hat H_{s}= -t\sum_{\sigma,i=1}^{N_{s}-1} \hat d_{i,\sigma}^{\dag} \hat
d_{i+1,\sigma}+\sum_{\sigma,\sigma^{^{\prime}},j,i=1}^{N_{s}}
 U_{i,j}^{\sigma,\sigma^{^{\prime}}} \hat
n_{i,\sigma}^{d} \hat n_{j,\sigma^{^{\prime}}}^{d}+
V_{o}\sum_{\sigma,i=1}^{N_{s}} \hat n_{i,\sigma}^{d} 
\end{equation}
\begin{equation}
\hat H_{i} = -t_{0}\sum_{\sigma}\hat c_{1,\sigma}^{\dag} \hat d_{1,\sigma} +
c.c. \nonumber
\end{equation}

where $\hat H_r$ corresponds to the ring, $\hat H_s$ to the stub,which is a
wire of finite length connected to the ring through the Hamiltonian $\hat H_i$
and $N_r$ and $N_s$ are the numbers of atomic sites belonging to the ring
and to the stub respectively. For the ring, it is assumed that $N_r+1=1$, and 
the magnetic flux is incorporated as the boundary condition (1).

We model the gate potential which controls
the state of charge of the  stub through the diagonal
elements of $\hat H_s$ given by $V_o$. For the case of ballistic transport in an
isolated ring, we have verified that
the properties of the system are weakly dependent upon the spatial range of
the e-e interaction. The effects of non-locality in the Coulomb interaction
could be restricted to the first neighbor intersite electronic repulion $U_1
$ that controls the effects that the charge accumulated in the stub has on
the currents circulating along the ring. Restricted to the intrasite and
first neighbour intersite contributions, the Coulomb interaction can be
written as $U_{i,j}^{\sigma,\sigma^{^{\prime}}}=
U_0\delta _{i,j}\delta _{\sigma ^{^{\prime }},-\sigma
}+U_1(\delta _{j,i-1}+\delta _{j,1+1})$.

The current is characterized by an energy scale that is given by the energy
difference between two successive states below the Fermi level ($E_f$).
For small values of $E_f$ it results to be:
\begin{equation}
\delta E=4\pi ^2t^2(2n+1)/N_r^2
\end{equation}
where $n$ is the integer number that defines the state wave vector $%
k=2\pi n/N_ra$, where $a$ is the lattice parameter. Although due to  
numerical limitations we are obliged to take a ring 
with small number of atoms, it is possible to get an adequate physical
representation of the system by scaling the Coulomb parameters $U_0$ and 
$U_1$ to the energy difference $\delta E $ defined at the neighbourhood of the Fermi
level. 

The size of a disordered
ring is an extremely important parameter to study the properties of
persistent currents because the localization length of the states near the
Fermi energy defines a characteristic length in the system. The behaviour is
different whether the size of the ring is bigger or smaller than this
magnitude. However, for an ordered ring, the absence of this length permits
to take an ideal small 1-D loop to study a real ring. Also it is necessary 
to suppose that it is thin enough as
to produce a sufficient energy separation of the different channels created by the lateral confinement.

To find the ground state we use a standard Lanczos algorithm. The
knowledge of it permits to calculate several quantities of interest for the
problem: the current in the ring, the charge in the stub and the total spin
of the ground state as function of the external magnetic flux and of the
gate potential $V_o$. The current is calculated as the mean value on the
ground state of the operator $\hat J$

\begin{equation}
\hat J= (4\pi et/h) {\rm Im}{(\sum_{\sigma ,i=1}^{N_r}(\hat c_{i,\sigma }^{\dag }\hat
c_{i+1,\sigma }-\hat c_{i+1,\sigma }^{\dag }\hat c_{o,\sigma }))}.
\end{equation}

\section{Results}

We study first an ordered ring just by assuming that $t_0=0$. The persistent
current of the non-interacting ring can be understood if we assume that the
one particle energy levels move along the free particle dispersion relation
as the magnetic field increases reducing the energy difference between the
levels that correspond to the wave vector $-2\pi i/N_ra$ and $2\pi(i+1)/N_ra$
for an arbitrary integer $i$. 

When the system possesses 2n particles there is an accidental degeneracy between the
state with $S=0$ and $S=1$ , where $S$ is the spin of the many body state.
This occurs for a magnetic flux $\phi _{*}=\phi _0/2$ for the case in which 
$n$ is an odd integer or when $\phi _{*}=0$ if $n$ is even. 
It is important to notice that the state with $S=1$ is the 
ground state only for this 
particular value of flux ($\phi _{*}$). If $\phi=\phi_* +\epsilon$, where 
$\epsilon$ is an arbitrary small number, the ground state is in the 
$S=0$ subspace. In this case there is degeneracy but not level crossing.

The Coulomb forces between the electrons shift the accidental degeneracy to
other values of $\phi_{*}$; $\phi _{*}<\phi _0/2$ for $n$ even and $\phi
_{*}>0$ for $n$ odd as it is shown in Fig. 1. It is straightforward to realize
that are the Coulomb forces which originates the level crossing.

When the number of particles is odd there is no accidental degener acy for
the non-interacting case. However, for interacting electrons the states with 
$S = 3/2$ and $S = 1/2$ could coincide in energy for $\phi = \phi_{*}$, $%
0<\phi_{*}<\phi_{0}/2$,  if the electronic repulsion were greater than the
kinetic energy difference between the last two occupied levels.

The two many body solutions mentioned above interchange their condition of
being the ground state and the first excited state at $\phi =\phi _{*}$.
If $U$ is small enough, a perturbative argument can be given as
follows. Let us suppose that the system has an even number of electrons and
that its ground state has $S=0$. Increasing the magnetic flux, when $\phi
>\phi _{*}$ the system could reduce its energy by occupying the state with
total spin $S=1$. Here the two electrons next to the Fermi level are unpaired:
one has a positive wave vector $k$ and the other a negative one and both the 
same value
of $ S_z$. However, for this process to
take place during an adiabatic evolution, the system has to go through an 
intermediate state in which one of the electrons hops from the state 
with pos itive $k$ to another with negative value of $k$ without flipping 
its spin. Due to the Coulomb interaction, the energy of the intermediate state 
is greater than the energy of the flipped spin final state $S=1$, and also is 
greater than the energy of the initial state S=0 due to its greater kinetic 
energy. So it defines the top of a potential barrier which separates both 
states. 
Under the hypotesis of thermodynamic equilibrium, at $\phi =\phi _{*}$ the
system changes its current discontinuously from a positive to a negative
value only if it is present an external spin flip mechanism capable 
of overcoming the potential barrier that exists separating both states in 
phase space. 

Numerically this adiabatic process can be simulated as follows. 
It is necessary to obtain the ground state of the system threaded by a flux $\phi_1=\phi_0 
+ \delta\phi$ where $\delta\phi$ is an arbitrary small flux. The wave function 
$|\alpha_{0}(\phi_{0})>$ corresponding to the ground state of the system under the 
effect of the flux $\phi_0$ is taken as the starting state to begin the process.
This state has a total spin $S_0$ and a total wave vector $K(\phi_0)$. It can 
be written as a linear combination of eigenstates of the 
Hamiltonian with flux $\phi_1$, $ |\alpha_{n}(\phi_{1})>$, 
\begin{equation}
|\alpha_{0}(\phi_{0})> = \sum_{n} a_n |\alpha_{n}(\phi_{1})> 
\end{equation}
Note that a change in the magnetic flux produce an spread of the state over 
the quantum number $K$ but not in the S quantum number.
Let us call the state that provides the dominant contribution to equation (7)  
by $|\alpha_{n_{0}}(\phi_{1})>$, which possesses the  
total wave vector $K(\phi_1)$ closest to the original $K(\phi_0)$, and the same
total spin ($S_0$) as the $|\alpha_{0}(\phi_{0})>$ state (within the numerical
precision). 

It is fundamental to know if in this process the system will remain in a unstable or 
metastable situation or not.
To clarify this point, let us define the three dimensional space $(S,K,E)$ 
where $S,K,E$ are  
the total spin ,the total wave vector and the energy respectively.
The behavior of the system will depend on the topology of this   
space in the neighbourhood of the starting point defined by the state
 $|\alpha_{0}(\phi_{0})>$.
  Suppose that the point corresponding to $S_0$ and $K(\phi_1)$ is an 
absolute minimum for the energy. Then, the system evolving from a small 
region around this point will go always towards it. 
Now, suppose that ther e is another minimum, which belongs to a different
(S,K) subspace, and that it is the true absolute minimum. The existence of
this minimum will modify the topology of the space around the 
starting point $S_{0},K(\phi_{0})$. However, if the energy difference between
the two minima is not great enough, the topology around a small vecinity 
of the starting point will remain unchanged. Then, the system evolving 
adiabatcally from this point will continue going toward  the $S_{0},K(\phi_{1})$
state.

An ideal tool to study this problem is the modified Lanczos alghorithm used to 
find the ground state  of the system. Ths method requires the definition of a new state $|\beta>$ by applying
the Hamiltonian to the $|\alpha_{0}(\phi_{0})>$ state and substracting the 
projection over it. 
The Hamiltonian represented in the basis ${|\alpha_{0}(\phi_{0})>,|\beta>}$ 
is diagonalized and the lowest energy state is taken as the starting new 
vector (as it is always a better aproximation to the real ground state).
This procedure is continued until convergence is reached. In the process of 
diagonalizing the 2x2 matrix at each step the lowest of the two diagonal  
elements ($d_\alpha,d_\beta$) determines the evolution of the initial state. 
The procedure follows a way in which
the energy obtained at the $n$ step is lower than the energy at the
step $n-1$. If the state $|\alpha_{n_{0}}(\phi_{1})>$ does correspond to 
a minimun in the parameter space the process goes toward it reducing the 
energy at each step. In the opposite situation, at an arbitrary step of the 
procedure the relation between  $d_\alpha$ and $d_\beta$ is inverted 
($d_\beta<d_\alpha$) and the system evolves to the orthogonal state $\beta$.
The quantum numbers corresponding to this state are in general arbitrary 
far apart from the quantum numbers of the original state even for $\delta\phi$
infinitesimally small. 
This is a procedure through which we are able to find numerically the frontiers
of the bistable region.

Then, when the magnetic flux is changed adiabatically in the neighborhood of $\phi
_{*}$, it is possible for the system to persist in a metastable state above
and below $\phi _{*}$, depending whether the flux is increasing or
decreasing. This introduces a hysteresis loop that appears as a bistability
in the I-$\phi $ characteristic curve as shown in Fig. 2 with a continuous
line.

In standard double-barrier eterostructures (DBH), transport involves non
linear phenomena reflected in the observation of multiple
stabilities in the I-V characteristic curve in the region where the device
exhibits a negative differential conduction \cite{7}. This property can be
thought to be produced  by an accumulation of  electronic charge in the well
at resonance and a rapid leakage of it when the applied voltage is just
taken the device out of resonance. This non-linear effect is essentially a
result of the interaction between the charges in the well and has been
extensively studied theoretically assuming that the potential profile seen
by the carriers, as they go along, depends in a self-consistent way upon the
charge distribution \cite{8}.

For our system, a bistability very similar to the one described above for
the DBH can be obtained in the $I-\phi$ characteristic curve. This is
because there is a part of the system, the stub in our case, the well for
the standard DBH, which is capable of acting as reservoirs of particles
controlled by the external applied potential for the case of the DBH or by
the external magnetic field or the gate potential in our case.

 Let us focus our attention to the ring weakly coupled to the stub taking 
  a small $t_{0}$. In order to discuss conceptually the bistable behaviour,  
we define the energy per particle $E_{\alpha}$
such that the total energy of the system is given by

\begin{equation}
E = \sum_{\alpha} E_{\alpha} \langle\hat n_{\alpha}\rangle 
\end{equation}

where

\begin{equation}
E_{\alpha} = e_{\alpha} + \sum_{i,j,\sigma,\sigma^{^{\prime}}} U_{i,j}^{\sigma,\sigma^{\prime}}
\langle\hat n_{i,\sigma}\hat n_{j,\sigma^{^{\prime}}}\rangle/N_{e} 
\end{equation}

and $\langle ... \rangle$ corresponds  to the mean value on the ground state
of the operators involved.
$e_{\alpha}$ refers to the energy of an electron without the Coulomb
interaction and $N_{e}$ is the number of electrons in the system.

In this case the phase space in which the bistability takes place is defined 
by the number of particles in the stub and the total spin .
The starting state to initialize the numerical calculation is assumed to be 
the solution that corresponds to the previous smaller (greater) magnetic 
flux or $V_0$.
This procedure simulates the behaviour of the system when the flux or the gate
potential is adiabatically increased (decreased). In Fig. 3 we present the 
variation of
the total energy and the $E_\alpha $ levels of the ring and the stub as a
function of the magnetic flux for the case with $U_0=0.3$, $U_1=U_0/2$, $t_0=0.0001$ and $V_0=0$ in units of the parameter $t$. For such a small value 
of the Coulomb parameters the pseudo single particle description is valid. Also
a small value of the $t_0$ parameter implies that one of the variables in the
parameter space ( the charge into the stub) becomes quasi discret.
The pseudo single particle description exhibits a
flux-nsensitive behaviour corresponding to a state localized at the stub,
and a flux depending state that describes a mobile particle inside the
ring. Due to a change in the magnetic flux, when a charge goes from one part
of the system to the other, a finit shift of the levels of each subsystem
occurs due to the Coulomb interaction together with a modification of the
total spin of the ground state. It is the dependence of these levels with the
state of charge, mainly in the stub but also in the ring, which drives the
bistb ility. The system can reduce its onergy at $\phi =\phi _{*}$ changing
its charge distribution, as shown in Fig.3. However, along an adiabatic
path as analyzed above, these two states of charge are separated in phase
space by a potential barrier which vanishes for a particular external
magnetic flux $\phi _*$ when the transition occurs. At these values of the
flux the energy of the total system changes discontinuously.

In this example, the bistability is unfavoured by the small value of 
the parameters $t_0,U_0$.
For greater values of $U_0$, in order to obtain an interchange of charge
between the ring and the stub, it is necessary to produce energy variations
that go beyond the range allowed by the magnetic flux. So, in this case the
gate potential is the relevant variable that controls the phenomenology of
the system. In Fig.4 the dependence of the current on $V_0$ is presented for 
$t_0=0.5t$.

For this value of $t_0$ the interchange of charge of the two subsystems permits
the existence of regions where the current in the ring is caused by a fractional
number of electrons.There are regions in which the charge is almost constant 
while in others it is highly dependent upon the gate potential. There, the
behaviour the total spin of the ground state can change as it occurs at 
$V_1$ and $V_2$ in Fig. 4. Associated with it a bistability appears, as discussed
above.

For a 1-D DBH the Coulomb blockade effect reflects that, as soon as
one electron enters the well region, the entrance of a second one is
excluded simply because it has to overcome the Coulomb repulsion produced by
the first electron already inside it. Here the Coulomb blockade
phenomenon is an intra-site effect. An electron cannot flow because another
one is already occupying the region through which it has to go to become an
electrical carrier. In our case, there is an inter-site Coulomb blockade
effect produced by the repulsion that the charge inside the stub exercises
over the flowing electrons within the ring in the nearby of the stub. This
is reflected in a reduction of the current in the regions where the charge
is stable, independent of $V_0$, in comparison with the current that
corresponds to a perfect ring with the same state of charge.

\section{Conclusions}

In conclusion we have presented a study of the many
body problem of an inhomogeneous closed ring enclosing a magnetic flux. 
We have shown that these systems have a bistable behaviour that can be associated to the 
physics of a non-linear system. We have also
studied the effect that the imperfection has upon the persistent current
flowing along the ring showing Coulomb blockade effect between the stub and
the ring. 
The development of sub-micrometer physics makes it possible to
construct this sort of devices. We hope the predicted phenomenology
could be a motivation for experimentalists to look for it in real systems.

\section{Acknowledgments}

One of us (E.V.A.) would like to acknowledge very interesting discussions
related to the subject of this work with M. Buttiker. We would like to acknowledge
the hospitality of the Department of Physics of UFF, Brazil, and the
Department of Physics of UBA, Argentina. This work was partially supported
by the Brazilian financial agencies CNPq and FINEP and by the grant
B-11487/4B005 of the Andes-Vitae-Antorchas Foundation. Also, G. Ch. would like
to acknowledge to UBA for a partial grant.

\newpage
\figure Fig.1:{(a) Two particles in a twelve sites ring. The energy is measured 
in unit Fof :he parameter $t$: (i) is the energy of the
S = 0 st.ate with $U_{0}$ = 0, (ii) is the energy of the S = 1 state and (iii) is
the energy of the S = 0 st$a$te with  $U_{0}= 0.5t$.
(b) Four particles in a ten sites ring : (i) is the energy of the
S = 0 state with $U_{0}= 0$, (ii) is the energy of the S = 0 state with $U_{0}
= 0.5t$, (iii) is the energy of the S = 1 state with $U_{0} = 0$ and (iv)is the
energy of the S = 1 state with $U_{0} = 0.5t$.} 
\figure Fig.2:{ (a) Current  versus flux for two particles in a twelve sites ring
and $U_0= 0.5t$: open squares correspond to the ground state, continuous line
correspond to the current of the metastable states.
(b) Current  versus flux for four particles in a ten sites ring
and $U_{0} = 0.5t$: open squares correspond to the ground state, continuous line
correspond to the current of the metastable states.
The current is measured in units of 4$\pi$$te/h$} 
\figure Fig3: {Four particles in a eight sites ring weakly coupled to a four sites stub
as shown in the right inset of a). The number of electrons in the
 ring is n.
a) The continuos line represents the energy of the  ground state. The crosses
 correspond to the energy obtained raising the flux adiabatically from 0. The open squares
represent the energy obtained decreasing the flux adiabaticaly from the right of $\phi_{2}$ .It is also shown in the lower inset of the figure.
(b) Pseudo single particle levels  of the system as a function of the magnetic 
flux. Open squares and triangles correspond to the first and second level
 of the  ring respectively.The continuous luine to the first level of the stub.
 The energy is measured in units of the parameter $t$ }
\figure Fig.4:{Four particles in a eight sites ring coupled to a four sites stub with
$t_{0}=0.5t$,$\phi = 0.4 \phi_{0}$ and $U_{0} = 4t$.  Open squares correspond
to a process where the gate potential is decreasing slowly from 0. Continuos
line corresponds to a situation in which the gate potential is increased 
adiabaticallyu from -6.
a)Current as function of gate potential.From $V_{2}$ to the left the open squares
correspond to the ground state. From $V_{2}$ to the right the ground staste
corresponds to the continuous line. Crosses represent the current of (n=1,2,3)
interacting particles in a perfect ring.
b) Charge in the stub (left axis) and total spin (rigth axis) as a function of
the gate potential. The current is measured in units of 4$\pi$$te/h$}

\end{document}